\documentclass[]{aa} 
\usepackage{graphicx}
\usepackage[varg]{txfonts}
\usepackage{natbib,twoopt}
\usepackage{breakurl}
\bibpunct{(}{)}{;}{a}{}{,} 
\makeatletter
 \newcommandtwoopt{\citeads}[3][][]{\href{http://adsabs.harvard.edu/abs/#3}%
 {\def\hyper@linkstart##1##2{}%
 \let\hyper@linkend\@empty\citealp[#1][#2]{#3}}}
 \newcommandtwoopt{\citepads}[3][][]{\href{http://adsabs.harvard.edu/abs/#3}%
 {\def\hyper@linkstart##1##2{}%
 \let\hyper@linkend\@empty\citep[#1][#2]{#3}}}
 \newcommandtwoopt{\citetads}[3][][]{\href{http://adsabs.harvard.edu/abs/#3}%
 {\def\hyper@linkstart##1##2{}%
 \let\hyper@linkend\@empty\citet[#1][#2]{#3}}}
 \newcommandtwoopt{\citeyearads}[3][][]%
 {\href{http://adsabs.harvard.edu/abs/#3}
 {\def\hyper@linkstart##1##2{}%
 \let\hyper@linkend\@empty\citeyear[#1][#2]{#3}}}
\makeatother

\begin{document}

   \title{Spectroscopic and dynamical properties of comet C/2018 F4, likely a true average former member of the Oort cloud
  \thanks{Based 
          on observations made with the GTC telescope, in the Spanish Observatorio del 
          Roque de los Muchachos of the Instituto de Astrof\'{\i}sica de Canarias, 
          under Director's Discretionary Time (program ID GTC2018-096).}}

   \author{J. Licandro\inst{1,2}
            \and
           C. de la Fuente Marcos\inst{3}
            \and 
           R. de la Fuente Marcos\inst{4}
            \and
            J. de Le\'on\inst{1,2}
            \and
           M. Serra-Ricart\inst{1,2}
            \and 
          A. Cabrera-Lavers\inst{5,1,2}
          }
   \offprints{J. Licandro, \email{jlicandr@iac.es}
             }

   \institute{Instituto de Astrof\'{\i}sica de Canarias (IAC), C/ V\'{\i}a L\'actea s/n, E-38205 La Laguna, Tenerife, Spain
                \and
              Departamento de Astrof\'{\i}sica, Universidad de La Laguna, E-38206 La Laguna, Tenerife, Spain
                \and
              Universidad Complutense de Madrid. Ciudad Universitaria, E-28040 Madrid, Spain
                \and
              AEGORA Research Group. Facultad de Ciencias Matem\'aticas, Universidad Complutense de Madrid. Ciudad Universitaria, E-28040 Madrid, Spain
                \and
              GRANTECAN, Cuesta de San Jos\'e s/n, E-38712 Bre\~na Baja, La Palma, Spain 
             }
     
   \date{Received 14 December 2018 / Accepted XX Xxxxxxxx XXXX}

 
  \abstract
  {The population of comets hosted by the Oort cloud is heterogeneous. Most studies in this area have focused on highly active
   objects, those with small perihelion distances or examples of objects with peculiar physical properties and/or unusual chemical 
   compositions. This may have produced a biased sample of Oort cloud comets in which the most common objects may be rare, particularly those with perihelia well beyond the orbit of the Earth. Within this context, the known 
   Oort cloud comets may not be representative of the full sample meaning that our current knowledge of the appearance of the average Oort cloud comet may not be accurate. Comet C/2018~F4 (PANSTARRS) is an object of interest in this regard.}
  {Here, we study the spectral properties in the visible region and the cometary activity of C/2018~F4, and we also explore its 
   orbital evolution with the aim of understanding its origin within the context of known minor bodies moving along nearly 
   parabolic or hyperbolic paths.}
  {We present observations obtained with the 10.4~m Gran Telescopio Canarias (GTC) that we use to derive the spectral 
   class and visible slope of C/2018~F4 as well as to characterise its level of cometary activity. Direct $N$-body simulations 
   are carried out to explore its orbital evolution.}
  {The absolute magnitude of C/2018~F4 is $H_{r} > 13.62\pm0.04$ which puts a strong limit on its diameter, $D < 10.4$~km, 
   assuming a $p_V = 0.04$ cometary-like value of the albedo. The object presents a conspicuous coma, with a level of activity comparable 
    to those of other comets observed at similar heliocentric distances. Comet C/2018~F4 has a visible spectrum consistent with 
   that of an X-type asteroid, and has a spectral slope $S'=4.0\pm$1.0~\%/1000~{\AA} and no evidence of hydration. The spectrum  
   matches those of well-studied primitive asteroids and comets. The analysis of its dynamical evolution prior to discovery 
   suggests that C/2018~F4 is not of extrasolar origin.}
  {Although the present-day heliocentric orbit of C/2018~F4 is slightly hyperbolic, both its observational properties and past 
   orbital evolution are consistent with those of a typical dynamically old comet with an origin in the Oort cloud.}
   

   \authorrunning{Licandro et al.}

   \titlerunning{GTC observations of C/2018~F4}
   
   \maketitle
%

\section{Introduction}

 In our solar system, a number of populations of small bodies are well studied and the notion of an average or typical member of such 
populations is well defined; good examples are the near-Earth objects or NEOs (see e.g. \citeads{2018Icar..312..181G}) or 
the Jupiter-family comets (see e.g. \citeads{1999A&A...352..327F} ; \citeads{2011MNRAS.414..458S}). Unfortunately, the appearance of an average Oort cloud comet remains unclear and this could be the result of most studies focusing on the extreme cases. 

The Oort cloud \citepads{1950BAN....11...91O} is a spherical structure that surrounds the solar system with an outer boundary 
located beyond 50\,000 to 200\,000~AU. The Oort cloud hosts a population of comets of heterogeneous nature (see e.g. 
\citeads{2001Natur.409..589S} ; \citeads{2003Icar..165..391G} ; \citeads{2004Icar..172..372F} ; \citeads{2016SciA....2E0038M} ; 
\citeads{2017AJ....154...53B}). Most authors consider the Oort cloud to have appeared very early in the history of the solar 
system, nearly 4.6~Gyr ago, and to be made of fossil debris from the primordial protoplanetary disc. Importantly, the solar 
system was born within a star cluster (see e.g. \citeads{2008Icar..197..221K}). Based on this information, 
\citetads{2010Sci...329..187L} suggest that perhaps over 90\% of the material currently present in the Oort cloud is of 
extrasolar origin, having been captured from the protoplanetary discs of other stars when the Sun was still part of the open star 
cluster or stellar association where it was born. An analysis of a sufficiently representative sample of objects from the Oort 
cloud should be able to either confirm or reject a dominant primordial extrasolar origin for the populations of small bodies 
hosted by the Oort cloud. Being able to clearly characterise the appearance of an average Oort cloud member may help in solving this 
difficult and important problem.

The current sample of known Oort cloud comets is likely biased in favour of relatively active objects, those with short perihelion 
distances, and those with unusual physical and/or chemical properties; unremarkable comets tend to be missing, perhaps neglected. 
Among the currently known minor bodies following nearly parabolic or hyperbolic paths  -- which may have their origin in the Oort 
cloud -- about 75\% have data-arcs spanning less than a month and consistently uncertain orbit determinations; only a small 
subsample has been studied spectroscopically. This suggests that our current perspective on the appearance of the average Oort cloud comet may be inaccurate. 
Comet C/2018~F4 (PANSTARRS) is an object of interest in this regard.

Comet C/2018~F4 was discovered by the Pan-STARRS survey  (\citeads{2004SPIE.5489...11K} ; \citeads{2013PASP..125..357D}) on 2018 
March 17 at 6.4~AU from the Sun and with an apparent magnitude $w$ of 20.4 (\citeads{2018MPEC....P...86G} ; 
\citeads{2018MPEC....H...21S} ; \citeads{2018MPEC....F..139T}). It was initially classified as a hyperbolic asteroid, A/2018~F4 
\citepads{2018MPEC....F..139T}, and was subsequently reclassified as a comet \citepads{2018MPEC....H...21S}. Its current heliocentric orbit 
determination is based on 185 data points, for an observation arc of 146 days, and it is hyperbolic at the 7$\sigma$ level, 
although the barycentric eccentricity is not hyperbolic (see Table~\ref{elements}) at almost the 21$\sigma$ level; $\sigma$ 
levels have been computed using the formal uncertainty on the eccentricity in Table~\ref{elements}. Its trajectory is 
approximately perpendicular to the plane of the solar system with the descending node being at about 6~AU from the Sun and the ascending 
node being far from any planetary path, stranded approximately midway between the orbits of Jupiter and Saturn.
%
%
     \begin{table}
        \fontsize{8}{12pt}\selectfont
        \tabcolsep 0.15truecm
        \caption{\label{elements}Heliocentric and barycentric orbital elements and 1$\sigma$ uncertainties of comet C/2018~F4 
                                 (PANSTARRS). 
                }
        \centering
        \begin{tabular}{lccc}
           \hline\hline
            Orbital parameter                                 &   & Heliocentric          & Barycentric \\
           \hline
            Perihelion distance, $q$ (AU)                     & = &   3.4417$\pm$0.0004   &   3.4355    \\
            Eccentricity, $e$                                 & = &   1.00077$\pm$0.00011 &   0.99771   \\
            Inclination, $i$ (\degr)                          & = &  78.160$\pm$0.003     &  78.256     \\
            Longitude of the ascending node, $\Omega$ (\degr) & = &  26.51923$\pm$0.00004 &  26.46807   \\
            Argument of perihelion, $\omega$ (\degr)          & = & 263.167$\pm$0.004     & 263.321     \\
            Mean anomaly, $M$ (\degr)                         & = &  -0.0020$\pm$0.0004   & 359.9899    \\
           \hline
        \end{tabular}
        \tablefoot{The orbit determination has been computed at epoch JD 2458227.5 that corresponds to 00:00:00.000 TDB, 
                   Barycentric Dynamical Time, on 2018 April 19, J2000.0 ecliptic and equinox. Source: JPL's SSDG SBDB.
                  }
     \end{table}
%
%

The aim of the research presented here is two-fold: (1) We aim to study the activity and surface properties of C/2018~F4 by obtaining a 
high-S/N image and a low-resolution spectrum in the visible, and  to compare them  to the activity and spectral properties observed in 
other comets (see e.g. \citeads{2018A&A...618A.170L}); and (2)  to explore its dynamics in order to determine whether it is an  Oort cloud comet, 
old or new, or perhaps an interstellar interloper. The paper is organised as follows. In Sect. \ref{sec:obs}, observations, data 
reduction, and spectral extraction are described. In Sect. \ref{sec:results}, we analyse the observed coma and the spectral 
properties of the comet, and compare  them to those of other comets. In Sect. \ref{sec:dynamics}, we explore the dynamical 
evolution of C/2018~F4 before finally presenting our conclusions in Sect. \ref{sec:conclusions}. 

\section{Observations \label{sec:obs}}

Images and low-resolution visible spectra of C/2018~F4 (PANSTARRS) were obtained in service mode on 2018 April 12 using the 
Optical System for Imaging and Low Resolution Integrated Spectroscopy (OSIRIS) camera spectrograph (\citeads{2000SPIE.4008..623C}
; \citeads{2010ASSP...14...15C}) at the 10.4~m Gran Telescopio Canarias (GTC). Two images, one with an exposure time of 180~s and 
the other of 30~s, were obtained between 0:33 and 0:43 UTC (at an airmass of 1.32) using the SLOAN {\em r'} filter. Three spectra, 
each one with an exposure time of 900~s, were obtained between 0:48 and 1:32 UTC (at an airmass of 1.29). OSIRIS has a mosaic of 
two Marconi 2048$\times$4096 pixel CCD detectors, with a total unvignetted field of view of 7.8$\times$7.8 arcminutes, and a 
plate scale of 0.127~"/px. In order to increase the S/N, we selected the 2$\times$2 binning and the standard 
operation mode with a readout speed of 200~kHz (with a gain of 0.95~e-/ADU and a readout noise of 4.5~e-). The tracking of the 
telescope matched the proper motion of the object during the observations. We found C/2018~F4 to be at 6.23 and 5.23~AU, heliocentric and 
geocentric distances, respectively, and its phase angle was $\alpha = 0\fdg9$ at the time of the observations. 

The spectra were obtained using the R300R grism in combination with a second-order spectral filter that produces a spectrum in the 
range 4800 to 9000~{\AA} with a dispersion of 32.25~\AA/px for the used 2.5" slit width. The slit was oriented in parallactic 
angle to account for possible variable seeing conditions and to minimise losses due to atmospheric dispersion. The three consecutive 
spectra were shifted in the slit direction by 10" to better correct for fringing. In addition, two G2V stars -- SA102-1081 and 
SA107-998 -- from the Landolt catalogue \citepads{1992AJ....104..340L} were observed immediately before and after the object, and 
at similar airmass (1.26 and 1.27, respectively) using the same spectral configuration. These stars are used as solar analogues 
to correct for telluric absorptions and to obtain relative reflectance spectra.  

\begin{figure}
 \centering
  \includegraphics[width=\linewidth]{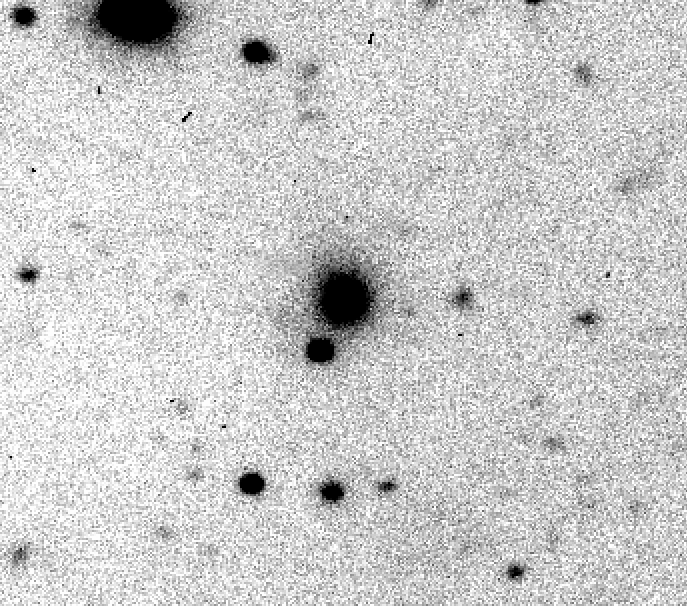}
  \caption{Image of C/2018~F4 (PANSTARRS) obtained on 2018 April 12. The image is a 90$\times$90" field; north is up, east to 
           the left. The object is found at the centre of the image and presents a faint coma indicative of some modest comet-like 
           activity.}
  \label{images}
\end{figure}

\begin{figure}
 \centering
  \includegraphics[width=\linewidth]{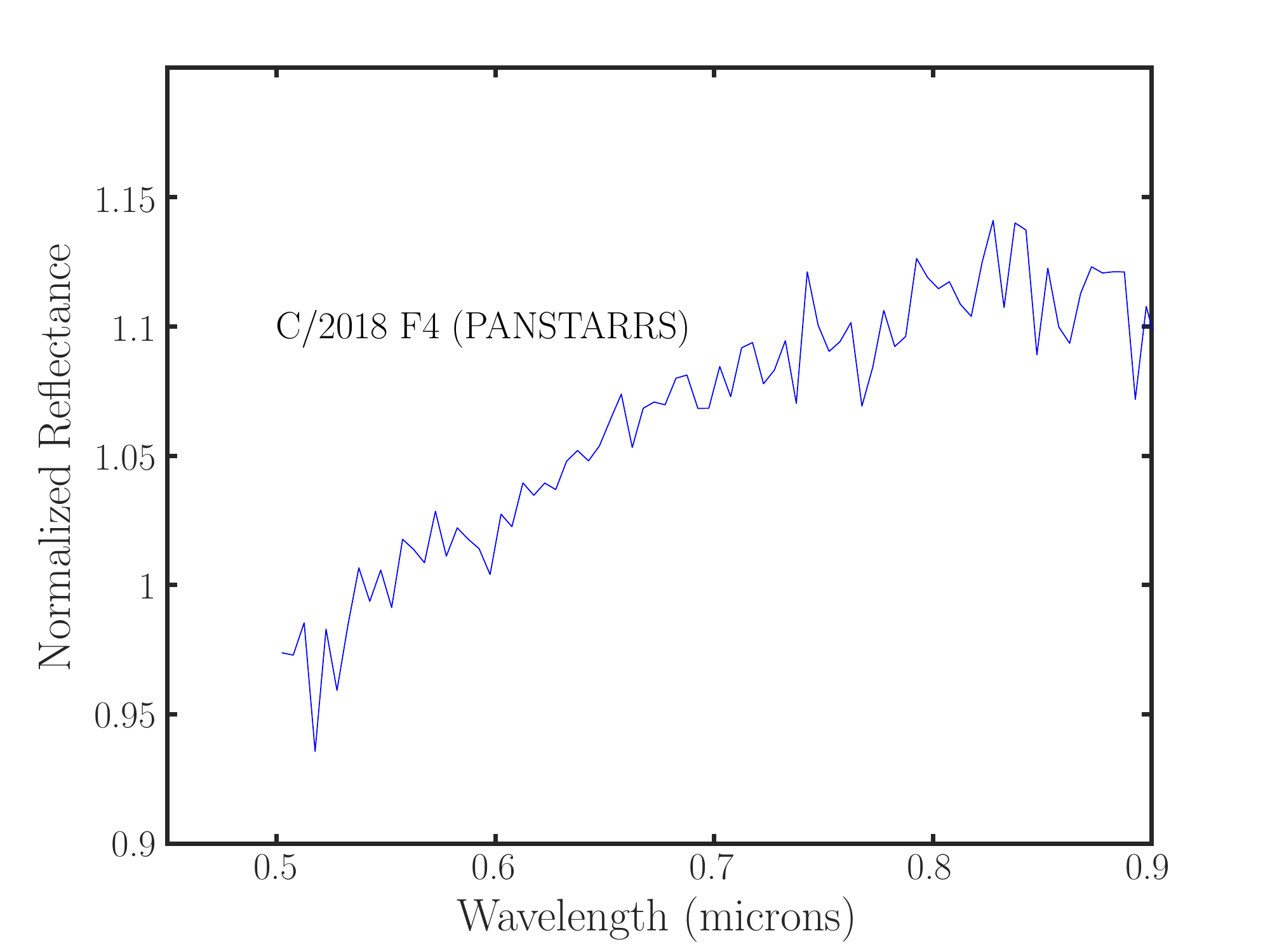}
  \caption{Visible reflectance spectrum of C/2018~F4 (PANSTARRS), normalized to unity at 0.55~$\mu$m.}
  \label{spectrum}
\end{figure}

Data reduction was carried out using standard Image Reduction and Analysis Facility (IRAF\footnote{IRAF is 
distributed by the National Optical Astronomy Observatory, which is operated by the Association of Universities for Research in 
Astronomy, Inc., under cooperative agreement with the National Science Foundation.}) procedures. The {\em r'} images 
obtained were over-scan and bias corrected, flat-field corrected using sky-flats, and flux calibrated using standard stars 
observed the same night. The 180~s image is shown in Fig.~\ref{images}. Spectral images were over-scan and bias corrected, and then 
flat-field corrected using lamp flats. The 2D spectra were extracted, sky background subtracted, and collapsed to one 
dimension. The wavelength calibration was done using Xe+Ne+HgAr lamps. Finally, the three spectra of the object were averaged to 
obtain the final spectrum. As pointed out above, two G2V stars were observed under the same conditions in order to improve the 
quality of the final comet reflectance spectra and to minimise potential variations in spectral slope introduced by the use of 
just one star. The averaged spectrum of the object was divided by that of each G2V star, and the resulting spectra were normalized 
to unity at 0.55~$\mu$m to obtain the reflectance spectrum. The final reflectance spectrum of C/2018~F4, binned to a resolution 
of 50~{\AA,} is shown in Fig.~\ref{spectrum}.
%
%
%
%
%
%
%
%

\section{Results and analysis \label{sec:results}}
In this section, we first analyse the images of C/2018~F4 (PANSTARRS) -- the deepest is shown in Fig.~\ref{images} -- to look for 
signs of comet-like activity, and measure the nuclear magnitude and activity levels. We then analyse its spectral properties, 
derive its taxonomical classification,  and compute its visible spectral slope.  Following the procedure described by \citetads{2018Icar..311...35D}, we also look for the typical 0.7~$\mu$m absorption band associated with the presence of hydrated minerals and observed on the surface of  some primitive asteroids. We finally compare these spectral properties  to those of known comet nuclei and dormant comets.

\subsection{Comet C/2018~F4 observed activity \label{sec:comaprop}} 

\begin{figure}
 \centering
  \includegraphics[width=\linewidth]{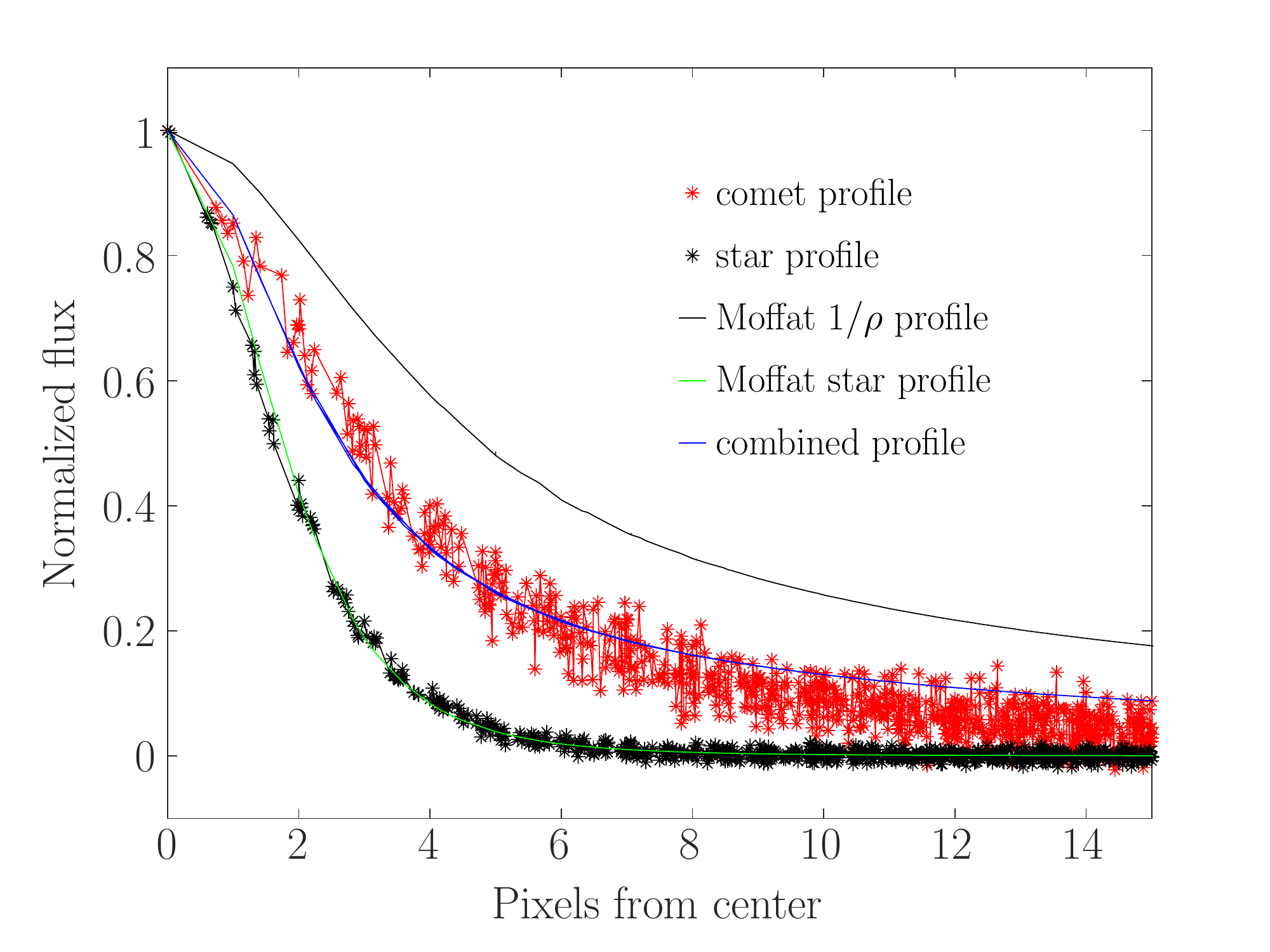}
  \caption{Normalized radial profile of C/2018~F4 (PANSTARRS) in the image shown in Fig.~\ref{images} (in red) compared to that 
           of a field star (in black). The Moffat profile of a point-like source (solid line in green) with the FWHM of the 
           stars in the images, and the corresponding profile of an extended object (solid line in black) with a $1/\rho$ profile 
           (the profile of an isotropic dust coma) are shown. The solid line in blue is the 50:50 linear combination of the 
           point-like source (nucleus) and $1/\rho$ profile (dust coma). The similarity of the combined profile to the observed 
           one suggests that the nucleus contributes significantly to the overall brightness of the comet.}
  \label{profiles}
\end{figure}

 Comet C/2018~F4 presents a rather obvious and compact faint coma as seen in Fig.~\ref{images}. The radial profiles of the 
comet and a field star shown in Fig.~\ref{profiles} clearly indicate that the point spread function (PSF) of the comet is wider 
(with a FWHM=6.4~pixels) than that of the field stars (FWHM=4.2~pixels). In order to assess the relative contributions of the 
nucleus and coma to the observed radial profile of the comet, we performed a simple analysis assuming an isotropic dust coma 
with surface brightness inversely proportional to the projected cometocentric distance ($\rho$). Using the IRAF task MKOBJECT, we created a synthetic image 
with two objects: (1) a point-like source and (2) an extended object with a $1/\rho$ profile. We assumed a Moffat PSF with the 
value of the FWHM measured for the field stars in the image of the comet. The radial profiles of these objects are also shown in 
Fig.~\ref{profiles}. We note that the point-like source (labelled as Moffat star profile) fits the profile of the stars in 
the comet image very well, while the $1/\rho$ profile (labelled as Moffat $1/\rho$ profile) does not fit that of the comet at all. A 50:50 
linear combination of the star and $1/\rho$ profiles matches the observed profile of the comet reasonably well, which strongly 
suggests that the contribution of the nucleus to the total brightness close to the optocentre of the comet is indeed significant.

 The apparent magnitude of the comet was measured using several apertures: $r'=21.22\pm0.04$ (6 pixels), $r'=19.90\pm0.04$ (12 pixels), and $r'=19.47\pm0.04$ (18 pixels). The brightness of the comet greatly increases  by 1.75~magnitudes when moving from 18 pixel to 6 pixel apertures. On the other hand, the magnitude variation between these two apertures computed for the field stars is only of 0.20. Such a difference is due 
to the contribution of the observed coma.  This clearly shows that the brightness of the comet nucleus, even using the small 
6 pixel aperture, is strongly contaminated by the coma as we also showed in the analysis of the profiles. An apparent magnitude 
$r'=21.22\pm0.04$ is simply a lower limit for the value of the nuclear magnitude of C/2018~F4, which in turn puts a robust limit 
to the nuclear magnitude and the size of the comet. In the combined profile described above, the coma contribution is 3.7 times larger 
than that of the nucleus, and thus the nuclear magnitude could be $\sim1.4$ magnitudes fainter. From the apparent magnitude, we 
derived an absolute nuclear magnitude of $H_{r} > 13.62\pm0.04$ using the procedure described in \citetads{2000Icar..147..161L}. 
Considering the solar colour transformations, the absolute magnitude in the visible is $H_{V} > 14.02\pm0.04$, and assuming an 
albedo of $p_V = 0.04,$ typical of comet nuclei (see \citeads{2018A&A...618A.170L}), this absolute magnitude limit corresponds to a 
diameter of $D < 10.4$~km for  comet C/2018~F4.
%
%
%
%

\begin{figure}
 \centering
  \includegraphics[width=\linewidth]{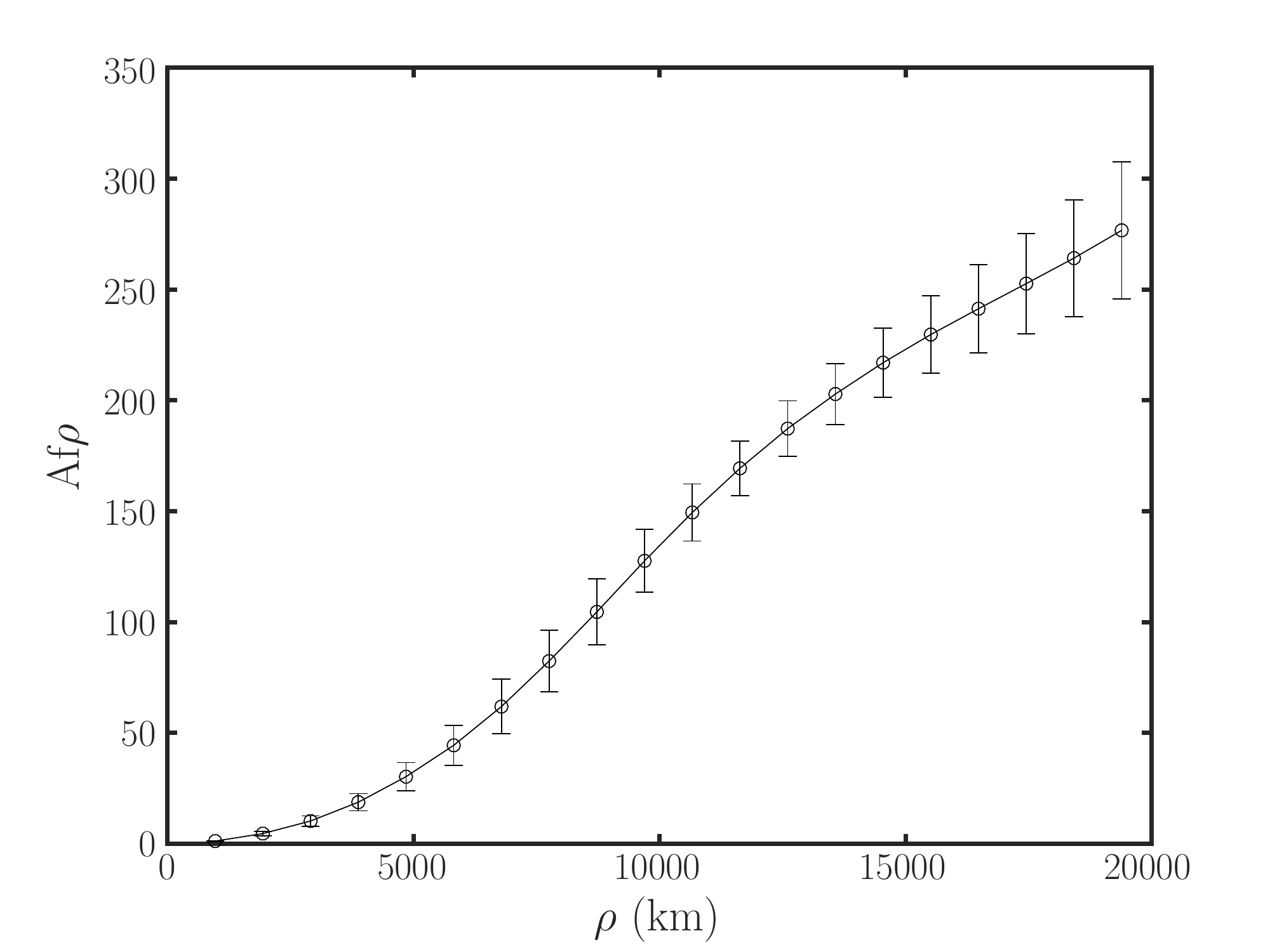}
  \caption{Variation of $Af\rho$ with $\rho$.}
  \label{Afrho}
\end{figure}

In order to evaluate the overall level of cometary activity (i.e. dust production rate) present during the observations, we 
computed the $Af\rho$ parameter -- or product between the albedo, the filling factor, and the radius of the coma 
\citepads{1984AJ.....89..579A} -- for different cometocentric distances ($\rho$) as shown in Fig.~\ref{Afrho}. At 
$\rho=10\,000$~km, $Af\rho = 148\pm13$~cm. Within the context of comet-like activity at large heliocentric distances, this value 
of $Af\rho$ agrees well  with those of P/2008~CL94 (Lemmon) and P/2011~S1 (Gibbs), of 106$\pm$3~cm and 76$\pm$8~cm, respectively 
\citepads{2016Icar..271..314K} and it is slightly below the mean $Af\rho$ value reported for comets observed at similar 
heliocentric distances by \citetads{2014A&A...561A...6M}, but still compatible with the less-active comets reported in this paper. 
The existence of comet-like activity beyond the zone of water-ice sublimation is very well known, and our results show that 
C/2018~F4 behaves in a similar manner to other comets observed at similar heliocentric distances.
%
%
%
%

\subsection{C/2018 F4 spectral properties  \label{sec:specprop}} 
                
A spectral slope of $S' = 4.0  \pm 1.0 $ \%/1000{\AA} is computed for the spectrum of C/2018~F4 following the $S'$ definition in 
\citetads{1996AJ....111..499L}, and considering the 0.55--0.86~$\mu$m wavelength range (where the observed reflectance 
is well represented by a linear fit). The spectrum is normalized to unity at 0.55~$\mu$m. The quoted uncertainty in the value of $S'$ 
has been computed as the standard deviation ($\sigma$) of the $S'$ values obtained for each single reflectance spectrum of the object; that is, the 
reflectance spectrum obtained for each single 900~s exposure time spectrum of the object and each single spectrum of the G2V stars 
obtained during the night when the observations were completed. The derived uncertainty ($\sim$1.0 \% /1000\AA) is compatible 
with the values (1--0.5 \%/1000\AA) usually obtained under good observational conditions, when several solar analogue stars have 
been observed. We  adopt this value as the  error of the computed slope instead of using the error obtained from the linear fit, which 
is much smaller. Using the online tool for modelling the spectra of asteroids, M4AST\footnote{\url{http://m4ast.imcce.fr/}} 
\citepads{2012A&A...544A.130P}, we obtained the taxonomical classification of C/2018~F4. Its spectrum corresponds to that of an 
X-type asteroid (\citeads{2004AJ....128.3070C} ; \citeads{2011Icar..214..131F}). As described by \citetads{2012A&A...544A.130P}, 
M4AST first applies a polynomial fit to the asteroid spectrum, with varying order, and then compares this fit at the corresponding 
wavelengths to templates of each taxonomical class defined by \citetads{2002Icar..158..146B} taxonomy. It then selects the taxonomic 
class with the smallest chi-squared value. 

Only a few visible or near-infrared  spectra of comet nuclei have been published, but all of them are featureless with a red slope 
in the 0.5 to 2.5~$\mu$m region, typical of X- or D-type asteroids and similar to the spectrum of C/2018~F4 presented here (see 
\citeads{2018A&A...618A.170L} and references therein). The photometric colours of comets are also typical of X- or D-type asteroids 
(see e.g. \citeads{2015AJ....150..201J}). 

The presence of aqueously altered minerals on asteroid surfaces can be inferred by a shallow spectral absorption band centred at 
0.7~$\mu$m. No signatures of this feature are present in the spectrum of C/2018~F4 (see Fig.~\ref{spectrum}). In any case, the 
absence of this feature does not mean that there are no hydrated minerals on the surface of the object. Several asteroids with 
hydrated minerals on their surfaces, inferred by a strong absorption feature in the 3-$\mu$m region, do not present the 0.7~$\mu$m 
band. In contrast, whenever the 0.7~$\mu$m band is present, the 3~$\mu$m band is also observed (see e.g. 
\citeads{2011epsc.conf..637H} ; \citeads{2015aste.book...65R}). The lack of evidence of water hydration on the surface of 
C/2018~F4 from its visible spectrum is also compatible with a cometary origin. Comet 67P/Churyumov--Gerasimenko is devoid of 
hydrated minerals \citepads{2017MNRAS.469S.712B} and the visible spectra of cometary nuclei do not present the features produced 
by hydrated minerals: the band centred at 0.7~$\mu$m and that at 0.5~$\mu$m. In particular, the visible spectrum of 
C/2018~F4 is different from those of the so-called Manx comets, long-period comets displaying only residual activity even at 
small perihelion distances \citepads{2016SciA....2E0038M}, as no significant dip beyond 0.75~$\mu$m is observed. 

 The results presented above should be taken with some caution, since the observed spectrum of C/2018~F4 is not exactly that 
of the comet nucleus; there is an important contribution of the coma ($\sim$ 60\% of the flux in the slit corresponds to scattered 
light from the coma according to the profile analysis presented above). In terms of slope determination and spectral 
classification, this should not affect our conclusions given the fact that the colour of the comet coma is similar to that of the 
nucleus (see \citeads{2015AJ....150..201J}). In contrast, such a contribution could have masked the presence of a weak absorption 
band like the 0.7~$\mu$m one due to aqueously altered minerals.

\section{Dynamics \label{sec:dynamics}} 

Aiming at exploring the dynamical evolution of C/2018~F4 (PANSTARRS), we have used data -- heliocentric and barycentric orbital 
elements and their uncertainties -- provided by Jet Propulsion Laboratory's Solar System Dynamics Group Small-Body Database (JPL's 
SSDG SBDB, \citeads{2015IAUGA..2256293G}).\footnote{\url{https://ssd.jpl.nasa.gov/sbdb.cgi}} Here, the full $N$-body calculations 
required to investigate the pre- and post-perihelion trajectories of this and other objects have been carried out as described by 
\citetads{2012MNRAS.427..728D} and do not include non-gravitational forces.  The orbit determination in Table~\ref{elements} did 
not require non-gravitational terms to fit the available astrometry; therefore, any contribution due to asymmetric outgassing 
is probably a second-order effect. Neglecting the role of non-gravitational forces in this case is unlikely to have any major 
impact on our conclusions. When nominal orbits are not used, the Cartesian state vectors are generated by applying the Monte Carlo 
using the Covariance Matrix method described by \citetads{2015MNRAS.453.1288D} and modified here to make it work with hyperbolic 
orbits; the covariance matrices necessary to generate initial positions and velocities have been obtained from JPL's 
\textsc{horizons},\footnote{\url{https://ssd.jpl.nasa.gov/?horizons}} which is also the source of other input data required to 
perform the calculations such as barycentric Cartesian state vectors for planets and other solar system bodies. This approach has previously 
been used to independently confirm that C/2017~K2 (PANSTARRS) is a bound and dynamically old Oort cloud comet 
\citepads{2018RNAAS...2b..10D}.

The analysis of the pre-perihelion trajectory of C/2018~F4 might shed some light on its true origin because although its 
present-day heliocentric orbital determination is nominally hyperbolic (see Table~\ref{elements}), it may or may not have followed 
an elliptical path in the past; realistic $N$-body simulations can help in investigating this critical issue. We have performed 
integrations backward in time of 1024 control orbits of this object; our physical model includes the perturbations by the eight 
major planets, the Moon, the barycentre of the Pluto--Charon system, and the three largest asteroids. A statistical analysis of 
the results indicates that about 38\% of the control orbits are compatible with the object coming from the interstellar medium at
low relative velocity with respect to the Sun. For this hyperbolic subsample and 1~Myr into the past, the average distance to the 
comet from the Sun was 0.3$\pm$0.2~pc (or 60\,855~AU), moving inwards at $-$0.5$\pm$0.4~km~s$^{-1}$ and projected towards 
$\alpha=16^{\rm h}~59^{\rm m}~12^{\rm s}$ and $\delta=+75\degr~18\arcmin~10\arcsec$ (255\degr$\pm$13\degr, +75\degr$\pm$2\degr) in 
the constellation of Ursa Minor (geocentric radiant or antapex) with Galactic coordinates $l=107\fdg41$, $b=+33\fdg18$, and 
ecliptic coordinates $\lambda=112\fdg56$, $\beta=+80\fdg02$, thus well separated from the ecliptic and the Galactic disc. The study 
of its post-perihelion trajectory requires the analysis of a similar set of $N$-body simulations, but forward in time; it will 
reach perihelion on 2019 December 4. Out of 1024 control orbits and after 1~Myr of simulated time, we observe that nearly 51\% 
lead the then unbound object towards interstellar space. 

 In order to better understand the past, present, and future orbital evolution of C/2018~F4 within the context of other 
objects with similar osculating orbital elements, we have searched JPL's SSDG SBDB and found that the heliocentric orbit 
determination in Table~\ref{elements} is somewhat similar in terms of perihelion distance, $q$, and inclination, $i$, to those of 
the long-period comets C/1997~BA6 (Spacewatch), $q= 3.436$~AU, $e = 0.999$, $i = 72$\fdg7, and C/2007~M2 (Catalina), $q = 3.541$~AU, 
$e = 0.999$, $i = 80$\fdg9, but also to those of the slightly hyperbolic comets C/1987~W3 (Jensen--Shoemaker), $q = 3.333$~AU, $e = 1.005$
(its barycentric eccentricity is also slightly hyperbolic, 1.000053), $i = 76$\fdg7, and C/2000~SV74 (LINEAR), $q = 3.542$~AU, 
$e = 1.005$ (as in the case of C/2018~F4, its barycentric eccentricity is not hyperbolic, 0.99994), $i = 75$\fdg2. In principle, these 
objects have not been selected to argue for some sort of physical or dynamical association with C/2018~F4, but to compare orbital 
evolutions of objects with similar values of $q$, $e,$ and $i$. However, it is true that the locations of the orbital poles 
($(L_{\rm p}, B_{\rm p}) = (\Omega-90\degr, 90\degr-i)$; see for example \citepads{1999ssd..book.....M}) of C/2000~SV74 and C/2018~F4 
are close in the sky, $(294\fdg2, 13\fdg8)$ versus $(296\fdg5, 11\fdg8)$, respectively. A small angular separation between orbital 
poles is indicative of a fairly consistent direction of the orbital angular momentum, which suggests that the objects are 
experiencing a similar background perturbation. In this context, C/2000~SV74 and C/2018~F4 may share the same dynamics even if 
they are not physically related -- their arguments of perihelion are nearly 180{\degr} apart. 

\begin{figure}
   \centering
   \includegraphics[width=\linewidth]{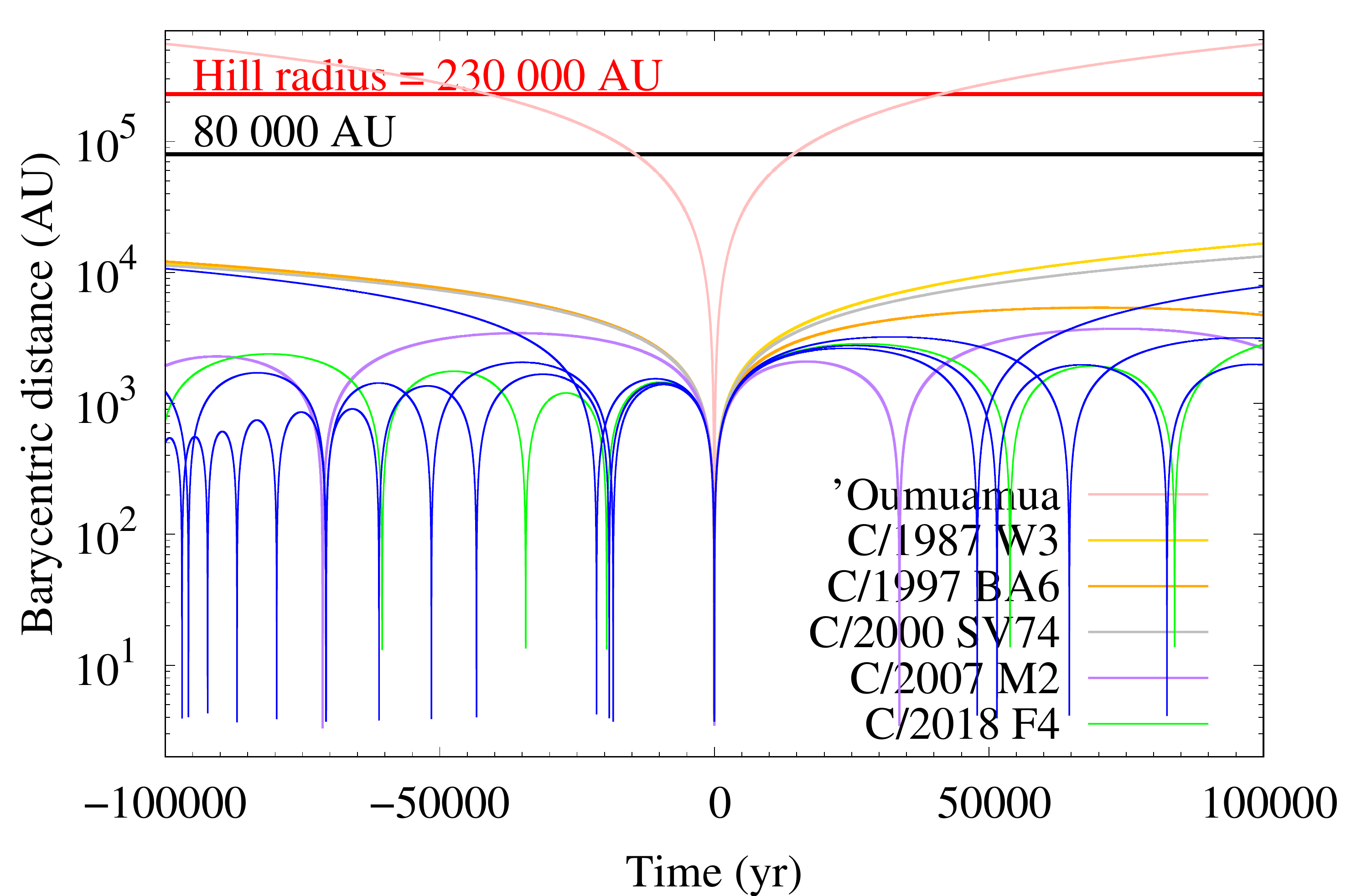}
   \includegraphics[width=\linewidth]{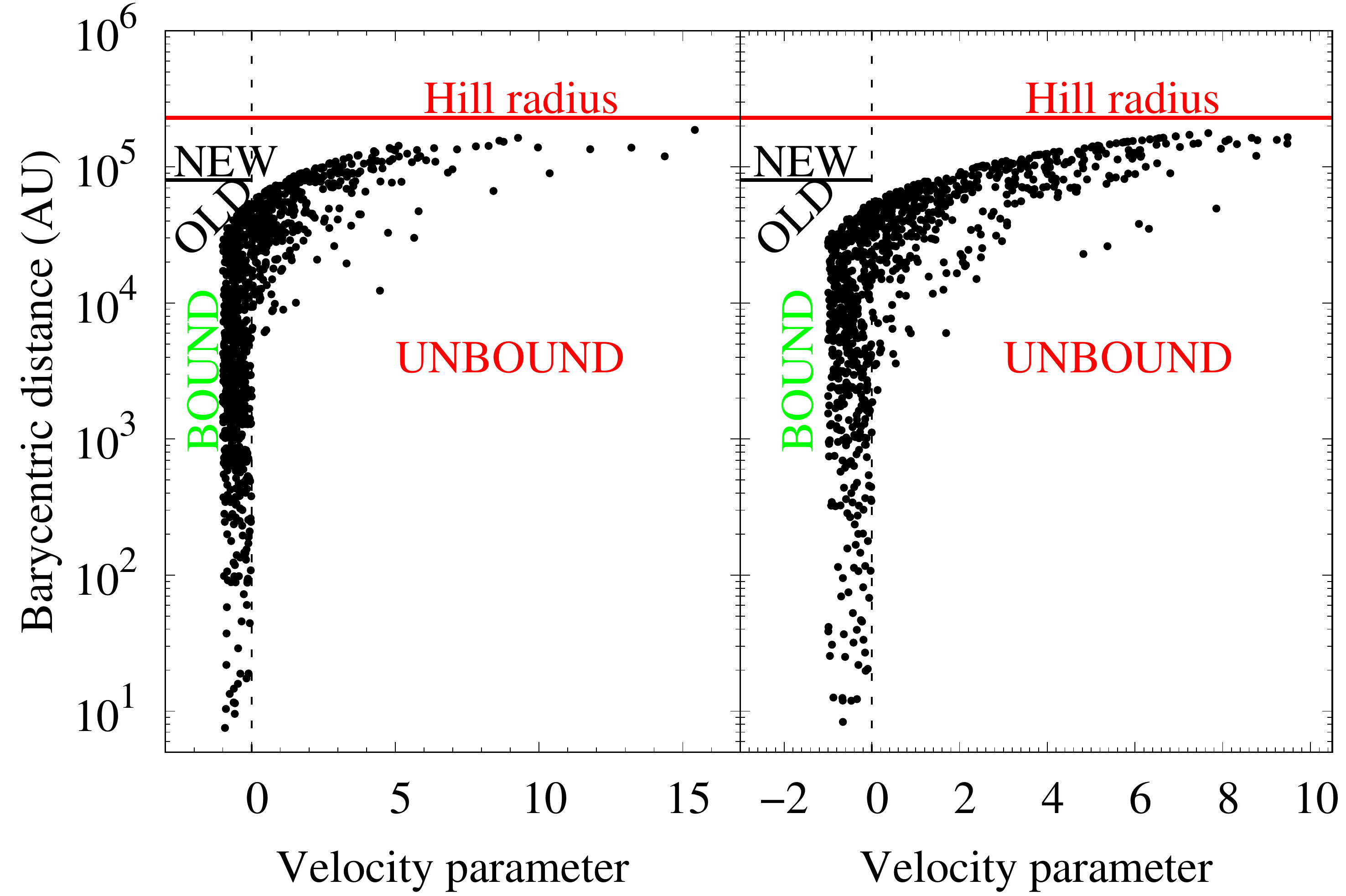}
   \caption{Evolution of the barycentric distance of 1I/2017~U1 (`Oumuamua), C/1987~W3 (Jensen--Shoemaker), C/1997~BA6 (Spacewatch), 
            C/2000~SV74 (LINEAR), C/2007~M2 (Catalina), C/2018~F4 (PANSTARRS) -- all nominal orbits -- , and a few representative 
            control orbits of C/2018~F4 (in blue) based on the current orbit determination (top panel). The zero instant of time, 
            epoch JDTDB~2458600.5, 27-April-2019, and only the time interval ($-$100\,000, 100\,000)~yr are displayed. Values of the 
            barycentric distance as a function of the velocity parameter 1~Myr into the past (left-hand side bottom panel) and 
            1~Myr into the future (right-hand side bottom panel) for 1024 control orbits of C/2018~F4. 
   \label{fig:1}}
\end{figure}

The top panel of Figure~\ref{fig:1} shows the (past and future) short-term orbital evolution of C/2018~F4 (in green, nominal orbit in 
Table~\ref{elements}) and those of a few representative control orbits (in blue). In addition, we show those of the  nominal orbits 
of 1I/2017~U1 (`Oumuamua), C/1987~W3, C/1997~BA6, C/2000~SV74, and C/2007~M2. The black line marks the aphelion distance -- $a \ 
(1 + e)$, limiting case $e=1$, semi-major axis, $a$ -- that signals the upper boundary of the domain of dynamically old Oort cloud 
comets (i.e. $a<40\,000$~AU, see \citeads{2017MNRAS.472.4634K}) as opposed to those that may be dynamically new, i.e. {\it bona 
fide} first-time visitors from the Oort cloud; the red line corresponds to the value of the radius of the Hill sphere of the solar 
system (see e.g. \citeads{1965SvA.....8..787C}). The bottom panels of Figure~\ref{fig:1} show the barycentric distance as a function 
of the velocity parameter 1~Myr into the past (left) and into the future (right) for 
1024 control orbits of C/2018~F4; the velocity parameter is the difference between the barycentric and escape velocities at the 
computed barycentric distance in units of the escape velocity. Positive values of the velocity parameter are associated with 
control orbits that could have been followed by putative visitors from outside the solar system (bottom-left panel) or 
that lead to ejections from the solar system (bottom-right panel). In summary, our $N$-body simulations and statistical 
analyses suggest that C/2018~F4 may be a dynamically old Oort cloud comet with a probability of 0.62 or, less likely, an interstellar 
interloper with a probability of 0.38. Given the fact that the inbound velocity may have been as low as 0.5$\pm$0.4~km~s$^{-1}$ 
, which is inconsistent with the lower limit for interstellar interlopers determined statistically by 
\citetads{2018MNRAS.476L...1D}, C/2018~F4 probably originated within the Oort cloud. In addition, 
C/1987~W3, C/1997~BA6, C/2000~SV74, and C/2007~M2 seem to be dynamically old comets.
%
%
%
%

\section{Conclusions \label{sec:conclusions}} 
In this paper, we have present observations of C/2018~F4 (PANSTARRS) obtained with GTC that we have used to derive the spectral 
class and visible slope of this minor body as well as to characterise its overall level of cometary activity. Direct $N$-body 
simulations were carried out to explore its orbital evolution. The object was originally selected to perform this study 
because early determinations of its orbital elements suggested a possible interstellar origin. Our conclusions can be summarised 
as follows.
\begin{enumerate}[(i)]
   \item We have determined a lower limit for the absolute magnitude of C/2018~F4, $H_{r} > 13.62\pm0.04$, and an upper limit of the diameter 
         $D < 10.4$km.
   \item We show that C/2018~F4 has a visible spectrum consistent with that of an X-type asteroid, with an spectral slope 
         $S'=4.0\pm1.0$~\%/1000{\AA} and no signs of hydrated altered minerals. This is consistent with the spectrum of a comet 
         nucleus.
   \item We show that the PSF of C/2018~F4 is definitely non-stellar and we confirm the existence of a detectable level of 
         comet-like activity when C/2018~F4 was observed at 6.23~AU from the Sun. We obtained an $Af\rho = 148\pm13$~cm measured 
         at $\rho=10\,000$~km, a value slightly below the mean $Af\rho$ value of comets observed at similar heliocentric distances 
         \citepads{2014A&A...561A...6M}, but still compatible with the level of activity shown by other distant comets.
   \item  The results of the analysis of an extensive set of $N$-body simulations indicate that the probability of C/2018~F4
         being a dynamically old Oort cloud comet is about 62\%.
   \item Conversely, the probability of C/2018~F4 having entered the solar system from interstellar space during the past 1~Myr is 
         about 38\% with an inbound velocity as low as 0.5$\pm$0.4~km~s$^{-1}$, inconsistent with the one expected for a true 
         interstellar interloper.
   \item The current path followed by C/2018~F4 is unstable, the probability of being ejected out of the solar system during
         the next 1~Myr is slightly above 50\%.  
\end{enumerate}

Based on our observational and numerical results, we favour an origin in the solar system for C/2018~F4. C/2018~F4 is likely a true representative of the average Oort cloud comet population. 

 
\begin{acknowledgements}
   The authors thank the referee A. Fitzsimmons for a constructive and useful report. 
   J. Licandro, M. Serra-Ricart, and J. de Le\'on acknowledge support from the AYA2015-67772-R  (MINECO, Spain). JdL acknowledges 
   support from from MINECO under the 2015 Severo Ochoa Program SEV-2015-0548.
   RdlFM and CdlFM thank S.~J. Aarseth for providing the code used in this research and  A.~I. G\'omez de Castro for providing
   access to computing facilities. This work was partially supported by the Spanish `Ministerio de Econom\'{\i}a y
      Competitividad' (MINECO) under grant ESP2015-68908-R. In preparation of this Letter, we made use of the NASA Astrophysics
      Data System, the ASTRO-PH e-print server, and the MPC data server.
   Based on observations made with the Gran Telescopio Canarias (GTC) installed in the Spanish Observatorio del Roque de los 
   Muchachos of the Instituto de Astrof\'{\i}sica de Canarias, in the island of La Palma, under Director's Discretionary Time 
   (program ID GTC2018-096).
\end{acknowledgements}

\bibliographystyle{aa} 
\bibliography{refsC2018F4.bib}

\end{document}